\documentclass[11pt,a4paper]{article}
\usepackage{mathrsfs}
\usepackage{amsmath}
\usepackage{amssymb}
\usepackage{dsfont}
\usepackage{esint}
\usepackage{amsmath,amssymb,amsfonts,a4wide,graphicx,bm,times,psfrag,wrapfig,sidecap}
\usepackage{cite}
\usepackage[colorlinks=true,linkcolor=black, citecolor=black,
urlcolor=black]{hyperref} 
\numberwithin{equation}{section}
\makeatletter \let\old@startsection=\@startsection
\renewcommand{\@startsection}[6]
{\old@startsection{#1}{#2}{#3}{#4}{#5}{#6\mathversion{bold}}}
\makeatother
\def\O{\Omega}
\def\Res{ \text{Res}}\def\Det{ \text{Det}}

\newcommand\re[1]{({\ref{#1}})}
\def\be{\begin{eqnarray}  }
    \def\ee{\end{eqnarray}}

\def\IZ{{\mathbb{Z}}}

    \def\ID{{\mathbb{D}}}

\def\ii{\varepsilon}
    \def\no{\nonumber}
    \def\la{\label}

\def\({\left(} \def\){\right)} 
\def\<{\langle} 
\def\>{\rangle} 
\def\[{\left[}
 \def\]{\right]} 
\def\tr{{\rm   tr} }
    \def\hf{ {\textstyle{1\over 2}} }

\def\CB{{\cal B}} 
\def\CO{{ \mathcal{ O} }}

    \def\CK{{ \mathcal{ K} }}
    
     \def\L{\Lambda}
     \def\CC{ {\mathcal P}}
       \def\CP{ {\mathcal P}}
       
       \def\CN{{ \cal  N}}
         \def\CV{{ \cal  V}}

 \def\CF{{\cal F}}
 \def\CY{{\cal Y}}
  \def\p{\partial}
  \def\a{\alpha}
 \def\b{\beta}

  \def\vp{\varphi}
  \def\th{\theta}

\def\d{\delta}
\newcommand{\caA}{{\mathscr A}}

\def\ket{ | 0 \rangle}
\def\bra{ \langle 0 | }
\def\zz{ { \{ \bf{\sigma} \} }}
  \def\Vii{\CV_\ii }
  \def\Qii{ \CQ_\ii }
   
\newcommand\encadremath[1]{\vbox{\hrule\hbox{\vrule\kern8pt
\vbox{\kern8pt \hbox{$\displaystyle #1$}\kern8pt}
\kern8pt\vrule}\hrule}} \def\enca#1{\vbox{\hrule\hbox{
\vrule\kern8pt\vbox{\kern8pt \hbox{$\displaystyle #1$} \kern8pt}
\kern8pt\vrule}\hrule}}

 \def\eff{{\text{eff}}}
  \usepackage{bm}% bold math
\def\O{\Omega}

\def\ee{\end{eqnarray}}  
\def\IZ{{\mathbb{Z}}}  
    \def\no{\nonumber} \def\la{\label} 
\def\({\left(} \def\){\right)}   \def\[{\left[} \def\]{\right]} \def\tr{{\rm tr} }
\def\hf{ {\textstyle{1\over 2}} } 
  \def\CB{{\cal B}} \def\CO{{
\mathcal{ O} }}  
\def\CK{{ \mathcal{ K} }} 
\def\L{\Lambda}
     \def\CC{ {\mathcal C}}  \def\CN{{ \cal N}}
       
      \def\CF{{\cal F}} \def\CY{{\cal Y}}
     \def\p{\partial} \def\a{\alpha} \def\b{\beta} 
        \def\vp{\varphi}
     \def\th{\theta}  
     
   \def\Li{ \text{Li}_2}
  \def\d{\delta}

\def\zz{ { { \bf  z} }}
\def\ff{ { { \bf  f} }}

\def\uu{ { {\bf u} }}

\def\vv{ { \bf v}}
\def\ww{ { \bf w }}

\def\vv { { \bf v  }}

  \def\k{\kappa}

   \def\k{\kappa}

 \def\slow{ \text{slow}}\def\fast{ \text{ fast}}

\begin{document}

\thispagestyle{empty}

\begin{flushright}
  %IPhT/t13/???  
\end{flushright}

\vspace{1cm}
\setcounter{footnote}{0}

\begin{center}

{\Large\bf   Semi-classical analysis of the inner product of Bethe states}

\vspace{20mm} 

Eldad Bettelheim$^{\ast}$ and Ivan Kostov$^{\star}$\footnote{\it
Associate member of the Institute for Nuclear Research and Nuclear
Energy, Bulgarian Academy of Sciences, 72 Tsarigradsko Chauss\'ee,
1784 Sofia, Bulgaria} \\[7mm]

{\it $^{\ast}$ Racah Inst.  of Physics, \\Edmund J. Safra Campus,
Hebrew University of Jerusalem,\\ Jerusalem, Israel 91904 \\[5mm]

{\it $^{\star}$ Institut de Physique Th\'eorique, CNRS-URA 2306 \\
             C.E.A.-Saclay, \\
             F-91191 Gif-sur-Yvette, France} \\[5mm]}

\end{center}

\vskip9mm

\vskip18mm

\noindent{ We study the inner product of two Bethe states, one of
which is taken on-shell, in an inhomogeneous XXX chain in the
Sutherland limit, where the number of magnons is comparable with the
length $L$ of the chain and the magnon rapidities arrange in a small
number of macroscopically large Bethe strings.  The leading order in
the large $L$ limit is known to be expressed through a contour
integral of a dilogarithm.  Here we derive the sub-leading term.  Our
analysis is based on a new contour-integral representation of the
inner product in terms of a Fredholm determinant.  We give two
derivations of the sub-leading term.  Besides a direct derivation by
solving a Riemann-Hilbert problem, we give a less rigorous, but more
intuitive derivation by field-theoretical methods.  For that we
represent the Fredholm determinant as an expectation value in a Fock
space of chiral fermions and then bosonize.  We construct a collective
field for the bosonized theory, the short wave-length part of which
may be evaluated exactly, while the long wave-length part is amenable
to a $1/L$ expansion.  Our treatment thus results in a systematic
$1/L$ expansion of structure factors within the Sutherland limit.  }

\newpage
\setcounter{footnote}{0}

 \section{Introduction and summary}
 
The computation of structure factors, matrix elements of operators
between eigenstates, analytically in exactly integrable systems
remains a challenging task.  In very small systems one may obtain
results by employing determinant formulas derived from the algebraic
Bethe ansatz.  The determinants are of matrices whose size increase
with the number of particles, such that fully analytical computations
go quickly out of hand.  In the thermodynamical limit of large number
of particles, the computation of such determinants becomes
intractable, except in special limits, usually accompanied by a
phenomenon which in physical terms may be viewed as a condensation of
excitations.
  
The most familiar cases are the condensation of magnons into bound
complexes with large spin in the Heisenberg ferromagnet as discovered
by Sutherland \cite{PhysRevLett.74.816} (hence the limit is sometimes
called the `Sutherland limit'), the condensation of solitons in the
quantum Sine-Gordon model to quasi-periodic solutions of the KdV
\cite{Babelon:Bernard:Smirnov:Quantization:Solitons} equation, or the
condensation of Cooper pairs in a superconductor
\cite{2011PhRvB..84v4503G}, to form either single or multiple
condensates, the latter being described by the Richardson model (a
particular example of Gaudin magnets).

More recently, bound complexes of magnons has been studied in the
context of the integrability in gauge and string theories
\cite{Beisert:2003xu}\cite{Kazakov:2004qf} (see also the review
\cite{Beisert-Rev}).  Some correlation functions in supersymmetric
Yang-Mills theories can be expressed in terms of inner products of
Bethe states in a chain of spins \cite{EGSV} and can be cast in the
form of a determinant \cite{Foda:3ptdeterminant}.  The thermodynamical
limit here is the limit of `heavy' fields in the Yang-Mills theory,
which correspond, by the AdS/CFT duality, to classical solutions of
the string-theory sigma model.  The three-point function of heavy
fields is exponentially small and can be thought of as a process of
semi-classical tunelling \cite{GSV}.

The leading order computations performed in \cite{GSV, 3pf-prl, SL}
gave an explicit expression of the exponent as a contour integral of a
dilogarithm.  In the present paper we give a method to compute the
higher orders of the semi-classical expansion and give an explicit
formula for the pre-exponential factor.

We focus on the XXX spin chain (the isotropic Heiseberg magnet), where
the thermodynamical limit corresponding to long-wavelength excitations
above the ferromagnetic vacuum.  In view of the applications, we
consider the more general case of an inhomogeneous spin chain with
twisted periodic boundary condition.  We will consider $M$-magnon
Bethe states in a chain of length $L$ in the thermodynamical limit
where $M, L\to\infty$ and $M /L\sim 1$.  Our goal is to propose a
systematic method for computing the $1/L$\ expansion for the
(logarithm of the) inner product of a Bethe eigenstate and an
off-shell Bethe state.
 
In particular we obtained an explicit expression for the subleading
term, given below.  Let $|\uu\rangle$ and $|\vv\rangle$ be two
$M$-magnon Bethe states in a XXX spin chain of length $L$,
characterized by the rapidities $\uu=\{u_1, \dots, u_M\}$ and
$\vv=\{v_1, \dots, v_M\}$.  One of the two states is required to be
on-shell in the sense that its rapidities satisfy the Bethe equations.
The two Bethe states are characterised by their pseudo-momenta
   \be p_\uu(x) = \sum_{j=1}^M {1\over x- u_j} - {L\over 2 x},\qquad
   p_\vv(x) = \sum_{j=1}^M {1\over x- v_j} - {L\over 2 x}.  \ee
In the semi-classical (thermodynamical) limit the root distributions
are described by continuous densities along one or several line
segments in the rapidity plane.  We call these line segments arcs
because of the typical form they take.  Each arc represents a branched
cut of the pseudo-momentum \cite{Beisert:2003xu, Kazakov:2004qf}.  The
inner product can be considered as the amplitude for semi-classical
tunelling with $\hbar = 1/L$ and as such is expected to have a $1/L$
expansion of the form
  \be \< \uu|\vv\>= e^{\CF_0+ \CF_1 + \dots},\qquad \CF_n\sim L^{1-n}.
  \ee
We obtained for the first two terms the following expressions in terms
of contour integrals:
  \be \la{CFzero} \CF_0&=& \oint \limits_{\CC} {dx\over 2\pi} \
  \Li[e^{ip_\uu(x)+ip_\vv(x)}]\, , \\
\no
\\
\la{CFone} \CF_1&=&- \hf \oint\limits_{ \CC\times\CC} {dx\, dy \over
(2\pi )^2}\ { \log\[ 1-e^{i p_\uu(x)+i p_\vv(x)} \] \ \log\[
1-e^{i p_\uu(y)+i p_\vv(y)} \]\over (x-y)^2} \, , 
\ee
where the contour of integration $\CC$ encircles the roots $\uu$ and
$\vv$.  This expression is valid, after redefinition of the
quasimomenta, for an inhomogeneous twisted XXX spin chain.  The first
term, containing a contour integral of the dilogarithmic function, is
of order $L$, because the typical size of the cuts is of order $L$.
It was first derived in \cite{GSV} for the special case when the
rapidities $\uu$ are sent to infinity, and for general $\uu$ and $\vv$
in \cite{3pf-prl, SL}.  The expression of the subleading term, which
is of order $L^0$, is the main result of this paper.
  
Our method is an improvement of the semi-classical computations in
\cite{3pf-prl, SL}, which used a representation of the Slavnov's
determinant \cite{NSlavnov1} in terms of a simpler quantity, the
$\caA$-functional\footnote{The $\caA$-functional generalises a
quantity defined in \cite{EGSV}, whose thermodynamical limit was
computed in \cite{GSV}.}.  The most symmetric form of such a
representation was found in \cite{sz}.  The present computation is
based on a new representation of the $\caA$-functional as a Fredholm
determinant, where the integration kernel is defined for a specific
contour in the complex plane.

We compute the semi-classical limit of this Fredholm determinant in
two different ways.  The first, rigorous, method consists in solving
the Riemann-Hilbert problem for the Fredholm kernel.  The second, less
rigorous but more intuitive, method uses field-theoretical formulation
of the Fredholm determinant in terms of free chiral fermions.  After
bosonization, we solve exactly the resulting field theory at small
distances to obtain an effective infrared field theory.  The
semiclassical expansion of the effective infrared theory can be also
thought of as Mayer expansion for a gas of dipole charges living on
certain contour in the rapidity plane.  The leading and the
sub-leading order are encoded in a saddle-point equation, which
resembles the `TBA-like' equations considered in
\cite{Nekrasov:2009aa}.

The paper is structured as follows.  In Section \ref{InnerProductXXX}
we recall the basics of the Algebraic Bethe Ansatz for the XXX spin
chain and the expression of the inner product in terms of the
$\caA$-functional.  In this section we also derive the determinantal
representation of the $\caA$-functional, which is the starting point
for our semi-classical analysis.  In Section \ref{RHApproach} we
develop the Riemann-Hilbert approach and find an explicit expression
for the subleading term.  In Section \ref{FieldTheory} we derive the
same result by field-theoretical methods.

\section{ Inner product   in  the inhomogeneous XXX$_{1/2}$ spin chain}
\la{InnerProductXXX}

\subsection{Algebraic Bethe Ansatz}

We first recollect some well known facts about the (twisted) periodic
XXX spin chain.  The inhomogeneous XXX spin chain of length $L$ is
defined by the monodromy matrix 
\be
\label{monodtheta}
M_\a(u)=\prod_{k=1}^L R_{\a k}(u-z_k ) \ee
where the auxiliary space is denoted by the index $\a$.  The rational
R-matrix can be taken in the form
\begin{equation}
\la{defRmatrix} R_{\a\b}(u)=\frac{u}{u+\ii}\,
I_{\a\b}+\frac{\ii}{u+\ii}\, P_{\a\b} ,
\end{equation}
with the operator $P_{\a\b}$ acting as a permutation of the spins in
the spaces $\a$ and $\b$.  The monodromy matrix depends of a set of
$L$ variables $\zz= \{ z_1, \dots, z_L\}$ called inhomogeneities,
associated with the sites of the chain.  Sometimes one uses the
notation
\be z_l = \th_l + \ii/2, \quad l=1,\dots, L. \ee
The isotropic Heisenberg Hamiltonian describes the homogeneous point
$\th_k=0$ or $z_k = \ii/2$.  The standard normalization of the
rapidity variable $u$ is such that $\ii = i$, but we prefer to keep
$\ii$ as a free parameter.  The monodromy matrix obeys the Yang-Baxter
equation
\begin{eqnarray}
\la{YBE} R_{\a\a'}(u-u') M _\a(u) M _{\a'}(u')= M _{\a'}(u')M
_\a(u)R_{\a\a'}(u-u').
\end{eqnarray}
Its diagonal matrix elements are traditionally denoted by
\be M _\a(u)=\left(\begin{array}{cc}A(u) & B(u) \\ C(u) &
D(u)\end{array}\right)_{\!  \a}.  \ee
The operators $A(u)$ and $D(u)$ act on the pseudo-vacuum $|\Omega
\rangle =|\uparrow \uparrow\ldots \uparrow\rangle$ as
\be \la{defad} A(u) |\Omega\rangle= a(u) |\Omega\rangle, \qquad D(u)
|\Omega\rangle= d(u) |\Omega\rangle, \ee
where the eigenvalues $a(u)$ and $d(u)$ are given, in the
normalization \re{defRmatrix} of the R-matrix, by
\begin{eqnarray}
 \la{normad} a(u)=1 \; , & \quad& d(u) = {Q_\zz(u)\over Q_\zz(u+\ii)}.
 \;
\end{eqnarray}
Here and below we will systematically denote by $Q_\ww$ the monic
polynomial with roots $\ww$:
\begin{align}Q_{\ww}(u) \equiv \prod_{i=1}^M (u-w_i) , \qquad
\ww\equiv \{ w_1, \dots, w_M\}\, .
\end{align}

Besides the inhomogeneities $\zz$ is convenient to introduce another
deformation parameter $\k$ by choosing twisted-periodic boundary
condition at length $L$.  The transfer matrix for the twisted chain,
\begin{equation}
T(u)=\tr_a \[ ( ^{1\ 0}_{0\ \k})M _a(u)\] =A(u)+\k\, D(u),
\end{equation}
commutes with itself for any value of the spectral parameter, and the
algebra of the matrix elements is the same as for the homogeneous XXX
model.

The Hilbert space is a Fock space spanned by states obtained from the
pseudo vacuum by acting with the `raising operators' $B(u)$:
\begin{eqnarray}
|\uu \rangle =B(u_1)\ldots B(u_M)|\Omega\rangle \;.
\end{eqnarray}
If the rapidities $\uu =\{u_1,\ldots,u_M\}$ are generic, the state is
called `off-shell', and the state is called `on-shell' if the
rapidities obey the Bethe Ansatz equations.  The Bethe equations for
the twisted chain read
\be 
\label{eq:BAEin}
{a(u_j)\over d(u_j)} + \k\, {Q_\uu( u_j+\ii)\over Q_\uu(u_j-\ii)}=1.
\ee
The `on-shell' states are eigenstates of the transfer matrix $T(x)$
with the eigenvalue
\begin{eqnarray}
t(x)=\frac{Q_\uu(x-\ii)}{Q_\uu(x)}+\k\,
\frac{d(x)}{a(x)}\frac{Q_\uu(x+\ii)}{Q_\uu(x)} \;.
\end{eqnarray}

\subsection{The inner product in terms of the $\caA$-functional}

We consider the bilinear form,
  \be (\vv, \uu) = \langle \O | \prod_{j=1}^M \CC(v_j)\ \prod_{j=1}^M
  \CB(u_j) |\O\rangle \ee
which we will refer to as inner product and which is related to the
scalar product by\footnote{This follows from the complex Hermitian
conjugation convention $B(u)^\dag = - C(u^*)$.}
    \be \(\uu,\vv \)= (-1)^M\< \uu^*|\vv\>.  \ee
The inner product of two Bethe vectors can be computed using the
commutation relations \re{YBE} and the action of the diagonal elements
of the monodromy matrices on the pseudo-vacuum \re{defad}.  The result
is written down by Korepin \cite{korepin-DWBC} as a double sum over
partitions.  It was shown by N. Slavnov \cite{NSlavnov1} that if one
of the two states is on-shell, the Korepin sum can be written as a
determinant.

The Slavnov determinant for a twisted periodic XXX chain can be
expressed in terms of a simpler quantity $\caA_\ww[f]$, which we call
$\caA$-functional, and which depends on the set of rapidities $\ww =
\{w_j\}_{j=1}^N$ and the function $f(u)$.  The inner product is equal,
up to a simple factor, to the $\caA$-functional \re{fDET} with $\ww=
\uu\cup \vv$ and $f (u) = \k\, d(u)/a(u)$
   \be \la{caAuv} \( \uu, \vv\) = \prod_{j =1}^M d(u_j )a(v_j )\ \caA
   _{\uu\cup \vv}[\k \, d/a]\, .  \ee
The formula (\ref{caAuv}) was derived by Y. Matsuo and one of the
authors \cite{sz} for purely periodic chain (no twist), but the proof
given there works without change also in the case of a
twist.\footnote{A twisted version of the determinant formula was also
discussed by Kazama, Komatsu and Nishimura \cite{Kazama:2013aa}.} This
form of the inner product is particularly useful due to its symmetry
in the rapidities $\uu$ and $\vv$.

The $\caA$-functional is defined as the ratio of $N\times N$
determinants \cite{SL}
\be \la{fDET} \caA_\ww[f]\equiv \det_{jk}\( w_j^{k-1} - f(w_j) \
(w_j+\ii)^{k-1}\)/ \det \( w_j^{k-1} \) \, .  \ee
Expanding the determinant, one obtains an alternative expression as a
sum over the partitions of $\ww= \{w_1,\dots, w_N\}$ into two disjoint
subsets $\ww_\a= \{ w_j\}_{j\in \a}$ and $\ww_{\bar\a}= \{
w_j\}_{j\in\bar \a}$\; :
\begin{eqnarray}
 \la{Aadefs}
\begin{aligned}
 \caA_{\ww} [f] &= \sum_{\a}\ \ (-1)^{|\a|} \prod _{ j\in\a} f(w_j)
 \prod_{j\in\a, k\in\bar\a} {w_{j}-w_k+ \ii \over w_{j} - w_k} .
\end{aligned}
\end{eqnarray} 
In this form the $\caA$-functional appeared (with $f= \k\, d/a$) as
one of the building blocks in the expression for the three-point
function in a supersymmetric Yang-Mills theory \cite{GSV}.

In this paper we will be interested in functional argument of the the
form
\be f(u) = \k\, {d(u)\over a(u)}=\k\, {Q_\zz(u)\over Q_\zz(u+\ii)}, \ee
which is relevant for the inhomogeneous twisted XXX chain.  We will
use a special notation for the $\caA$-functional as a function of the
magnon rapidities $\ww= \{ w_j\}_{j=1}^N$, the inhomogeneities $\zz=\{
z_l\}_{l=1}^L$, the twist $\k$ and the shift parameter $\ii$:
\be \caA_{\ww, \zz}^{[\ii, \k]} \equiv \caA_\ww[ \k\, d/a].  \ee
In these notations the inner product reads
   \be \la{caAuvuz} \( \uu, \vv\) = \prod_{j =1}^M d(u_j )a(v_j )\
   \caA^{[\ii, \k]} _{\ww, \zz } \qquad \qquad \qquad (\ww= \uu\cup
   \vv).  \ee

\subsection{ The $\caA$-functional as an $N\times N$ determinent}

\def\CQ{{\cal Q}}

In this paper we will use another determinant representation,
\begin{equation}
\label{detAK}
   \caA _{\uu, \zz}^{[\ii, \k]} = \det \(\mathds{ 1} -\kappa\, K \)
\end{equation}
where $\mathds{ 1}$ is the the $N\times N$ identity matrix and the
matrix $K$ has matrix elements
\be \la{defKmatr} K_{jk} &=&\frac{\ii E_j}{u_{j} -u_k+ \ii }
\label{DefK} \qquad\qquad (j,k=1,\dots, N)\, , \\
\la{defEj}
E_j &\equiv&   {Q_{\zz}(u_j)\over Q_{\zz}(u_j+\ii  )}
 \prod_{k(\ne j)} {u_{j} -u_k+ \ii    \over u_{j}-u_k}
 \, 
 .
 \label{EjDef}
\ee
To prove \re{detAK}, we write the sum over the partitions in
\re{Aadefs} as a double sum in one of the subsets:
\begin{eqnarray} 
\la{ExpAb}
\begin{aligned}
\caA_{\uu, \zz } ^{[\ii, \k]} &= \sum_{\a}\ (-\k )^{|\a|} \prod _{
j\in\a} { E_j } \ \prod _{ j, k\in\a\, ; \, j\ne k} {u_{j}-u_k\over
u_{j}-u_k+ \ii }
\end{aligned}
 \end{eqnarray}
and apply the Cauchy identity $(j,k\in\a)$
\begin{eqnarray}
\la{cauchy} \prod_{j\ne k} {u_{j}-u_k\over u_{j}-u_k+ \ii } = \det{
\ii \over u_{j}-u_k+ \ii } .
\end{eqnarray}
The new determinant representation \re{detAK} has the advantage that
it exponentiates in a simple way:
\begin{eqnarray}
\la{logDet} \log \caA _{\uu, \zz}^{[\ii, \k]}=- \sum_{n=1}^\infty\
{\k^n\over n} \sum_{j_1, \dots, j_n=1}^N {\ii\ E_{j_1}\over
u_{j_1}-u_{j_2}+\ii} \ {\ii\ E_{j_2}\over u_{j_1}-u_{j_3}+\ii} \
\cdots \ {\ii\ E_{j_n}\over u_{j_n}-u_{j_1}+\ii}.
\end{eqnarray}

\subsection{Semiclassical limit: from discrete data to meromorphic functions}

We are going to study the semi-classical limit $L\to\infty,
N\to\infty$ with $\a=N/L$ finite, when the roots $\uu$ arrange in one
or several arcs of macroscopic size.  We can also choose an
$L$-dependent normalisation of the rapidity variable so that $\ii\sim
1/L$.  Then the typical size of the arcs will be of order $L^0$.
  
For our task it is advantageous to replace the discrete data $\uu$ and
$\zz$ by the external potential
\be 
\la{DefPhi} \Phi(x) \equiv \log Q_\uu(x) -\log Q_\zz(x). 
\ee
In the semi-classical limit the arcs condense in one or more cuts of
the meromorphic function $p(u) \equiv \p\Phi(u)$.  The discontinuities
across the cuts are approximated by continuous densities which change
slowly at distances of order $\ii$.

The crucial observation which will allow to reformulate the problem in
terms of the external potential $\Phi$ is that factors $E_j$ defined
in \re{defEj} are the residues of the same meromorphic function at
$x=u_j$:
\begin{eqnarray}
    \la{EjQ} E_j= {1\over \ii} \ \underset{x\to u_j}\Res \Qii (x)\,
    \qquad (j=1,\dots, N) .
  \end{eqnarray}
The function $\Qii $ is defined as
\be \la{defQii} \Qii (x) = {Q_\uu(x+\ii)\over Q_\uu(x)} {Q_\zz(x)\over
Q_\zz(x+\ii)}= e^{ \Phi(x_j+\ii)-\Phi(x_j) }.  \ee
With the help of \re{EjQ} one can write the sum in the $n$-th term of
the series \re{logDet} by a multiple contour integral along a contour
$\CC_\uu$ which encircles all the roots $\uu$.

The weight function $\CQ_\ii(x) $ strongly fluctuates when $x$
approaches $\uu$ or $\zz$, but if $x$ is far from both $\zz$ and
$\uu$, it changes slowly at distances $\sim \ii$.  Our goal is to
reformulate the inner product in terms of contour integrals where the
contour of integration $\CC$ is placed far from the singularities of
the function $\CQ_\ii$, unlike the original contour $\CC_\uu$.  Then
the weight $\CQ_\ii$ can be replaced by
\be \la{QcalisQ} \CQ(x)= \lim_{\ii\to 0} \CQ_\ii(x)= e^{ \ii
\p\Phi(x)}.  \ee

We will achieve this goal by two different, and in a sense
complementary approaches.  The first one relies on the solution of a
Rieman-Hilbert problem, while the second one uses field-theoretical
concepts.  In both approaches the general idea is the same as in the
original computation in \cite{GSV}, namely to introduce a cutoff $\L$
such that $|\ii|\ll \L\ll L|\ii|$ and split the problem into a fast
(short-distance) and slow (large-distance) parts.  The final result
does not depend on the precise value of the cutoff $\L$.

\section{Riemann-Hilbert Approach}
\la{RHApproach}

We will represent the linear operator with matrix \re{defKmatr} as an
integral operator acting in a space of functions with given analytic
properties.  Then the determinant (\ref{detAK}) takes the form of a
Fredholm determinant.  In the semi-classical limit it is possible to
split the resolvent for the Fredholm kernel into slow and fast pieces.
The fast piece can be evaluated exactly, while the computation of the
slow piece is done by solving a standard scalar Riemann-Hilbert
problem.

\subsection{The $\caA$-functional as a Fredholm determinant  }

We represent an $N$-dimensional vector $\ff= \{f_1,\dots, f_N\}$ as a
meromorphic function $f(u) $, which has poles at $u=u_j$ with residues
$f_j$ and no other singularities:
\begin{align}
\label{calDefinition}
f(x)
\equiv \sum_j \frac{f_j}{x-u_j}.
\end{align}
The functions
\be \la{basis} e_j(x)
= {1\over x-u_j} \qquad  (j=1,\dots, N).
\ee
form a canonical basis in the $N$-dimensional space of meromorphic
functions analytic everywhere except on $\uu$.  The matrix
\re{defKmatr} defines a linear operator in this basis.

In order to be able to compute traces, we should also give a
functional representation of the dual space.  The elements $\tilde
\ff$ of the dual space with respect to the scalar product $\tilde \ff
\cdot \ff= \tilde f_1 f_1 +\dots + \tilde f_Nf_N$ can be mapped to the
space of functions $\tilde f(x) $ which are analytic in the vicinity
of $\uu$.  With such a function we associate a dual vector
$\tilde{\ff} $ with coordinates $ \tilde f_j\equiv \tilde f(u_j)$.
This function is of course not unique.  The scalar product may then be
represented by a contour integral
\begin{equation}
\< \tilde f |f \> = \oint\limits_{\CC_\uu} {dx\over 2\pi i} \ \tilde
f(x) f(x),
\end{equation}
where $\mathcal{C}_\uu$ is a contour surrounding the $\uu$'s and is
contained in the domain of analyticity of $\tilde f$.

We will use Dirac notations $ f(x) = \<x|f\>, \tilde f(x)= \<\tilde
f|x\>$, $f_j = \< j|f\>$ and $\tilde f_j= \< \tilde f|j\>$, so that
\begin{align}
|f\>=\sum_{j=1}^N |j\> \< j|f\>,\quad \< \tilde f| =\sum_{j=1}^N\ \<
\tilde f|j\rangle \<j| .
\end{align}  
The functional representations of the basis vectors in the direct and
the dual spaces are
 \be |j\>\to \<x|j\> \equiv {1\over x-u_j}; \qquad \< j|\to \< j |x\>
 : \ \ \< j | u_k\> = \d_{jk} \quad (k=1,\dots, N).  \ee
The functions corresponding to the elements of the dual basis are
defined up to an arbitrary meromorphic function that vanish on $\uu$.

The functional representation of the matrix $K$ is given by an
integral operator $\CK$, which acts on the function $f(x)\equiv \<
x|f\>$ as
\begin{align}\label{KactionQ}
\<x|\mathcal{K}|f\> =\frac{1}{2\pi i} \oint_{\CC_\uu} \Qii(y) \,
f(y+\ii ) \frac{dy}{x-y}\, ,
\end{align}
where the function $\Qii$ is defined by \re{defQii}.  The contour of
integration $\CC$ in this formula is chosen to encircles all $u_j$ but
leaves the points $z_l-\ii$ as well as the point $x$ outside.  Note
that there are no other poles of the integrand, as the poles of
$f(y+\ii)$ are compensated by the zeros of $\CQ_\ii$.  Applying
\re{EjQ}, we obtain the action in the canonical basis
  \be [\CK f]_j =\sum_k {\ii E_jf_{k}\over u_j-u_k+\ii},
  \ee 
 
 which agrees with (\ref{DefK}).  Now the $N\times N$ determinant
 \re{detAK} takes the form of a Fredholm determinant
\begin{equation}
\label{detAKF}
\caA _{\uu, \zz}^{[\ii, \k]}
      =
   \det \( 1 -\kappa\,  \CK
\)    .
\end{equation}

One may recast the definition of the Fredholm Kernel, Eq.
(\ref{KactionQ}),\ in the following operator form, which will be very
useful in extracting the semiclassical limit,
   \be \la{Kdecomp} \CK= \CP_\uu \ e^{-\Phi} \ \ID_\ii \, e^{\Phi},
   \ee
where $\ID_\ii =e^{\ii\p}$ is the shift operator, acting as $\ID_\ii
f(x) = f( x+\ii)$, and $\CP_\uu=\CP_\uu^2$ is the operator projecting
onto the space of functions \re{calDefinition}:
\be [\CP_\uu f](x) = \oint_{\CC } {du\over 2\pi i} {f(u)\over x-u} \
\qquad (u\ \text{is outside }\ \CC ).  \ee
Let us stress on the important fact that the contour $\CC$, which
encircles the set $\uu$, can be placed at macroscopic distance from
the roots in $\uu$.  Indeed, the resolvent has no other poles than
$x=u_j$ and $x=z_l-\ii$.  Along the deformed contour $\CC$ the factor
$\CQ_\ii$ in the Fredholm kernel changes slowly at distances of order
$\ii$.  We will denote functions in the image of $\mathcal{P}_\uu$
with a `$+$' subscript and functions in the kernel of
$\mathcal{P}_\uu$ with a `$-$' subscript.  Thus it will be implied
that
\be \la{defplus} \mathcal{P}_\uu \, g_+ =g_+, \quad \mathcal{P}_\uu\,
g_-=0.  \ee
We will also use the `$+$' subscript to denote functions which are in
the image of $\mathcal{P}_\uu$ up to a polynomial, namely
$\mathcal{P}_\uu g_+ = g_+ + P$, where $P$ is a polynomial.

\subsection{Resolvent of the Fredholm kernel}

We proceed by writing the logarithm of the $\mathscr{A}$-functional as
follows:
\begin{align}
\log\mathscr{A}_{\uu,\zz}^{[\ii ,\kappa]}
 =\int_0^\kappa \frac{d\alpha}{\alpha } \tr\left[ \mathds{1}
 -\left(\mathds{1}-\alpha K \right)^{-1} \right],
\label{IntegrateResolvent}
\end{align}
which leaves us with the task of computing the trace of the resolvent
$\left(\mathds{1}-\alpha K \right)^{-1}$.  Here $K$ is the $N\times N$
matrix defined in \re{defKmatr}.  We wish to find a functional
representation of the resolvent, which we denote by $\mathcal{F}$ and
define as
\begin{align}
\<x|\mathcal{F}|f\> = \sum_i \frac{\left[(\mathds{1}-\alpha
K)^{-1}\ff\right]_i}{x-u_i}.
\end{align}
One can compute $\tr(\mathds{1}-\alpha K)^{-1}$ in terms of
$\mathcal{F}$ as follows:
\begin{align}
\tr(\mathds{1}-\alpha K)^{-1} = \sum_{i=1}^N \ \< i|\mathcal{F}| i\> .
\label{TrInTildeIs}
\end{align}
The function 
\begin{align}
 F(x,u_i)\equiv
\<x|\mathcal{F}|i\>
\end{align}
will appear repeatedly in the following.  We shall analytically
continue $F(x,u_i)$ in the variable $u_i$, so that $u_i$ can be
thought of as a general complex variable rather than one of the roots
from the set $\uu$.  This analytical continuation is not unique, but
the ambiguity is arguably exponentially small and will be neglected in
the following.  To compute $F(x,u)$ explicitly, we note the following
identity:
\begin{align}
&\<x|(\mathds{1}-\alpha \mathcal{K})|f\> =e^{-\Phi(x)}\left(1-\alpha
\ID_\ii\right) e^{\Phi(x)} f(x)
+\alpha\sum_{l=1}^L \frac{ Q_{\uu}(z_l)}{Q'_{\zz}(z_l)}
\frac{e^{-\Phi(z_l-\ii )}f(z_l)}{x-z_l+\ii },
\label{oneplusalphaK}
\end{align}
which is obtained making use of \re{Kdecomp} and taking the projection
by removing the poles explicitly, the latter being located at the
points $z_l-\ii $.  The function $F(x,u)$ satisfies by definition
\begin{align}
\label{Fsatisfiesdiscr}
 & (1-\alpha \mathcal{K})F(x,u_j)= \frac{1}{x-u_j }\qquad (j=1,\dots, N).
 \end{align}
Analytically continuing both sides away from the set $\uu$ we obtain
for the meromorphic function $F(x,u)$ the equation
\begin{align}
\label{Fsatisfies}
 & (1-\alpha \mathcal{K})F(x,u)= \frac{1}{x-u }.
 \end{align}
Substituting the function $F(x,u) $ for $f(x)$ in Eq.
(\ref{oneplusalphaK}) leads to
 \begin{align}
 F(x,u) = e^{-\Phi(x)} \left(1-\alpha \ID_\ii\right)^{-1}e^{\Phi(x)}
 \left(\frac{1}{x-u}- \alpha\sum_{l=1}^L e^{-\Phi(z_l-\ii )}\frac{
 Q_{\uu}(z_l)}{Q'_{\zz}(z_l)}\frac{F(z_l,u)}{x-z_l+\ii }\right).
 \label{SelfConstCenter}
\end{align}
Then the equation may be solved self-consistently by treating
$F(z_l,u)$ on the right hand side as external parameters, solving for
$F(x,u)$ and then requiring that by evaluating $F(x,u)$ at $x=z_l$ we
recover these same parameters.  Indeed, setting $x$ to $z_l$ in Eq.
(\ref{SelfConstCenter}) and representing $\left(1-\alpha
\ID_\ii\right)^{-1}$ as $\sum_n\alpha^n \ID_\ii^n $, one realizes that
only the $n=0$ term in this sum contributes, which makes the
application of the self-consistency straightforward, leading to:
\begin{align}\left(\begin{array}{c}
F(z_1,u) \\
F(z_2,u) \\
. \\
. \\
F(z_L,u)
\end{array} \right)
=(\mathds{1} -  \tilde K)^{-1}\left( \begin{array}{c}
\frac{1}{z_1-u} \\ \frac{1}{z_2-u} \\ . \\ . \\ \frac{1}{z_L-u}\end{array}\right),
\label{SelfConsist}
 \end{align}
with the $L\times L$ matrix $\tilde K$ given by
 \begin{align}
 \tilde K_{ln} =- \frac{Q_{\zz}(z_{n}-\ii
 )Q_{\uu}(z_n)}{Q'_{\zz}(z_n)Q_{\uu}(z_n-\ii )}\ \frac{1}{z_l-z_n+\ii
 }.
 \end{align}

\subsection{Separation  into fast and slow pieces}

 We compute $\tr (1-\alpha K)^{-1}$ by splitting the rhs of Eq.
 \re{SelfConstCenter} into two parts, $F=F^{\fast}+F^{\slow}$, as
 follows:
\begin{align}
&F^\fast(x,u) =e^{-\Phi(x)} \left(1-\alpha \ID_\ii \right)^{-1}
e^{\Phi(x)}\frac{1}{x-u},\label{FLiDef} \\ 
&F^{\slow}(x,u) =- \alpha \, e^{-\Phi(x)} \left(1-\alpha
\ID_\ii\right)^{-1}e^{\Phi(x)}\sum_{l=1}^L e^{-\Phi(z_l-\ii )} \frac{
Q_{\uu}(z_l)}{Q'_{\zz}(z_l)}\frac{F(z_l,u)}{x-z_l+\ii }.
\label{FzDef}
\end{align}
We start by computing the contribution of $F^\fast(x,u)$ to $\tr
\left(\mathds{1}-\alpha K \right)^{-1}$:
\begin{align}
&\sum_{j=1}^N \underset{x\to u_j} {\rm Res} F^{\fast}(x,u_j)
=\sum_{j=1}^N \left(1+\sum_{n=1}^\infty \frac{\alpha^n}{n\ii }
e^{\Phi(u_j+n\ii )} \frac{Q_{\zz}(u_j)}{Q'_{\uu}(u_j)} \right) =
\nonumber\\
&=N- \oint\limits _\CC {dx\over 2\pi i}\ e^{-\Phi(x)}\log\left(1-\alpha\, 
 \ID_\ii\right) e^{\Phi(x)} 
.\label{FLi2}  
\end{align}

In order to find the contribution of $F^{\slow}$ to $\tr
\left(\mathds{1}-\alpha K \right)^{-1}$, we define an integral
operator $\mathcal{F}^{\slow}$ with the following action:
\begin{align}
 \<x|\mathcal{F}^{\, \slow}|f\> = \oint\limits_\CC {du\over 2\pi i} \
 F^{ \slow}(x,u)f(u).
\end{align} 
We are interested in computing the trace $\sum_{j=1}^N\, \< j
|\mathcal{F}^{\slow}|j\>,$ the contribution of $\mathcal{F}^{\,
\slow}$ to (\ref{TrInTildeIs}).  For that we introduce another
complete set of states $| m\>$, represented by functions $ \< x|m\> =
f_m(x) $ analytic on and inside the contour $\CC$, and a dual set $\<
m|$, obeying $\<m'|m\>=\delta_{m',m},$ represented by functions $\<
m|j\> = \tilde f_m(j),$ the domain of analyticity of which contains
$\CC$.  The quantum number $m$ is discrete if the contour $\CC$ is
compact and continuous otherwise.  For example, if there exists a
circle centered at the point $x_0$ such that the set $\uu$ is inside
the circle and the set $\zz$ is outside the circle, then we let
$\mathcal{C}$ be this circle and choose $f_m(x) = (x-x_0)^{-m}$,
$\tilde f_m(x) = (x-x_0)^{m-1}$ for $m\ge 1$.
 
The definition of $|m\>$ and $\<m|$ imply
\be \la{completeness} \oint_\CC {dx\over 2\pi i} \<
m|x\>\<x|m'\>=\d_{m,m'}\ , \qquad \sum_m \< x |m\>\<m|x'\>= {1\over
x-x'}\, , \ee
which allows to write the trace as
\begin{align}
\sum_{j=1}^N\< j|\mathcal{F}^{\slow}|j\> =& \sum_{j=1}^N
\sum_m\<j|\mathcal{F}^{\slow}|m\>\<m| j\> = \sum_{j,m}\<m|j\>\<
j|\mathcal{F}^{\slow}|m\> \nonumber\\
=&\sum_m \<m|\mathcal{F}^{\slow}|m\>.
\label{TrInMspace}
\end{align}   
The first equality follows from the relation $\< x|j\> =
\sum_m\<x|m\>\<m|j\>$ for any $x$, which is true by the definition of
$|m\>$ as a complete set.  To prove the last equality, we will show
that $\< x|\mathcal{F}^{\slow}|m\> $ has only simple poles at the
$u_i$'s and has additional singularities only around the $z_i$'s.  For
such functions, the sum of $|j\>\<j|$ acts as the identity operator,
and one can write
 \be \la{projectorI} \sum_{j=1}^N\<m|j\>\< j|\mathcal{F}^{\slow}|m\> =
 \oint\limits_\CC {dx\over 2\pi i} \< m|x\>\<x| \CF^\slow |m\>.  \ee
Indeed, the left hand side of \re{projectorI} is the sum of the $N$
residues of the integrand inside the contour $\CC$.  The last identity
\re{projectorI} of is then a consequence of \re{TrInMspace} and
\re{completeness}.

We are left with the task of showing the above-mentioned analytical
properties of $\< x|\mathcal{F}^{\slow}|m\> $.  Namely we must show
that $\< x|\mathcal{F}^{\slow}|m\> $ has only simple poles at the
$u_i$'s.  Indeed, combining (\ref{FzDef}) and (\ref{SelfConsist}), we
obtain
\begin{align}
\<x|\mathcal{F}^{\slow}|m\>&=\alpha e^{-\Phi(x)} \left(1-\alpha
\ID_\ii\right)^{-1}e^{\Phi(x)} \times \nonumber\\
\no
&\\
&\times \left(\begin{array}{c} \frac{ Q_{\zz}(z_1-\ii
)Q_{\uu}(z_1)}{Q_{\uu}(z_1-\ii )Q'_{\zz}(z_1)}\frac{1}{x-z_1+\ii } \\
. \\
. \\
  \frac{ Q_{\zz}(z_L-\ii )Q_{\uu}(z_L)}{Q_{\uu}(z_L-\ii
  )Q'_{\zz}(z_L)}\frac{1}{x-z_L+\ii } \end{array}\right)^t(\mathds{1}-
  \alpha \tilde K^t)^{-1}\left( \begin{array}{c}f_m(z_1) \\ .  \\ .
  \\ f_m(z_L)\end{array}\right),
\end{align} 
whereupon the required analytic properties become apparent.  This
concludes the proof of \re{TrInMspace}.

Writing (\ref{TrInMspace}) in terms of $F^{\slow}$ yields
\begin{align}
 \sum_i\< j | {\CF}^{\slow}|j\> &= \sum_m \oint {dx \over 2\pi i}\oint
 {du\over 2\pi i} \<m| x\> F^{\slow}(x,u) \< u| m\> = \nonumber\\
 &=   \oint  {dx \over 2\pi i}\oint  {du\over  2\pi i} \frac{F^{\slow}(x,u)}{u-x}  
  , \label{FzContribution} 
\end{align}
where the contour for the the integral in $u$ encircles the contour
for the integral in $x$.  Taking into account that $ \oint
\frac{F^{{\fast}}(x,u)}{u-x} du=0$, we can also write
\begin{align}
\sum_{j=1}^N\<j|\mathcal{F}^{\slow}|j\> = \oint {dx \over 2\pi i}\oint
{du\over 2\pi i} \ \frac{F(x,u)}{u-x} \label{TrFz} .
\end{align}
Combining (\ref{FLi2}), (\ref{IntegrateResolvent}) and (\ref{TrFz}),
we obtain:
\begin{align}
\label{GeneralMain}
 \log\mathscr{A}_{\uu,\zz}^{[\ii ,\kappa]} =\int_0^\kappa
 \frac{d\alpha}{\alpha} \left[\oint\limits_\CC \frac{dx}{2\pi
 i}e^{-\Phi(x)}\log\left(1-\alpha \ID_\ii \right) e^{\Phi(x)} +
 \oint\limits_\CC \frac{dx}{2\pi i}\oint\limits_\CC \frac{du}{2\pi i}
 \ \frac{F^{}(x,u)}{x-u} \right].
\end{align}
The representation (\ref{GeneralMain}) of $\log\mathscr{A}_{\uu,\zz}$
is only useful if the $1/N$\ expansion of $F$ is computable.  This
turns out to be the case, and we undertake the task of performing this
expansion in the following.

\subsection{Semi-classical  expansion of the slow piece }

The resolvent $F(x,u)$ satisfies the defining equation
(\ref{Fsatisfies}), which can be written, making use of \re{Kdecomp},
in the form
\begin{align}\label{RHFull}
\CP_\uu\left[F(x,u)-\alpha \CQ_\ii(u) F(x+\ii ,u) \right]=
\frac{1}{x-u}.
\end{align}
We can treat $\ii $ in the argument of $F$ as a small parameter.
Indeed, since the contour $\CC$ is at macroscopic distance from the
arcs formed by the roots $\uu$, the function $F(x, u)$\ changes slowly
at distances of order $\ii$.  We thus obtain the expansion
\begin{align}
\CP_\uu\left[F(x,u)-\alpha \CQ_\ii(x)\left( F(x,u) + \ii F'(x,u)
+\dots \right)\right]= \frac{1}{x-u}.
\end{align}
This equation is solved order by order in powers of $\ii $,
$F=F^{(0)}+F^{(1)}+\dots$, with the leading order satisfying
\begin{align}
\label{LeadingOrderResolvent}
\CP_\uu\left[\left(1-\alpha\CQ_\ii(x)\right) F^{(0)} (x,u)\right]=
\frac{1}{x-u} \, ,
\end{align}
while the next to leading order can be easily seen to be given by
\begin{align}\label{SubLeadingResolvent}
F^{(1)} (x,u)= -\alpha\ii \oint\limits_\CC {dv\over 2\pi i}\
F^{(0)}(x,v) \CQ (v) \partial_{v}F^{(0)}(v,u)\, ,
\end{align}
where $\CQ(x)$ is the limit as $\ii\to0$ of $\mathcal{Q}_\ii(x)$, Eq.
\re{QcalisQ}.  Computing yet higher orders is likewise mechanical.

The function $F^{(0)}$ satisfies a standard problem in the theory of
integral equations, Eq.  (\ref{LeadingOrderResolvent}).  There is a
standard method of solution of such equations \cite{Muskelishvili}.
Namely, we decompose $1-\alpha \CQ_\ii (x)$ into two parts,
\begin{align}
1-\alpha \CQ(x)= U_-(x)U_+(x),\label{Decomposition}
\end{align}
where $U_+(x)$ is analytic away from the arcs formed by the roots
$\uu$ and behaves at $x\to\infty$ as $U_+(x)\to x^n$ for some $n$ of
order $1$, while $U_-(x)$ is analytic around the arcs, having no zeros
around the arcs.  We give an explicit expression for $U_\pm$ in the
following.  We can write Eq.  (\ref{LeadingOrderResolvent}) as
\begin{align}
&\left(1-\alpha \CQ(x) \right)F^{(0)}+g_- = \frac{1}{x-u}
\end{align}
for some function $g_-$ regular around the arcs.  Using the
decomposition \re{Decomposition}, we write
\begin{align}
&U_+F^{(0)}+\frac{g_-}{U_-}  = \frac{1}{(x-u)U_-}.
\end{align}
We now apply the projector $\CP_\uu$ to both sides of this equation.
The second term on the left hand side drops out while the first term
yields a polynomial of degree $n-1$ in $x$ for $n$ positive, and zero
otherwise.  We denote this polynomial as $P_{n-1}(x;u)$ as it is also
a function of $u$.  We thus have
\begin{align}\label{PMminus1Def}
\CP_\uu[U_+(x)F^{(0)}(x,u)] =U_+(x)F^{(0)} (x,u)-P_{n-1}(x;u),
\end{align} from which one obtains:
\begin{align}
\la{F0U} F^{(0)}(x,u)=\frac{1}{U_+(x)} \left\{\CP_\uu\left[
\frac{1}{(x-u)U_-(x)}\right]+ P_{n-1}(x;u) \right\}.
\end{align}
Finally, using the analytical properties of $U_-$ and the definition
of $\CP_\uu$ it is easy to see that
\be
\CP_\uu\left[\frac{1}{U_-(x)(x-u)}\right]=\frac{1}{U_-(u)(x-u)},
\ee
and Eq.  \re{F0U} simplifies to
\begin{align}\label{F0Solution}
F^{(0)}(x,u)=\frac {P_{n-1}(x;u)}{U_+(x)}+\frac{1}{U_+(x)U_-(u)(x-u)}.
\end{align}

The coefficients of the polynomial $P_{n-1}(x,u)$ can be found by
solving (\ref{LeadingOrderResolvent}) for $F^{(0)}$ around infinity to
order $x^{-n}$ and using $P_{n-1}(x;u) =
\left[U_+(x)F^{(0)}(x,u)\right]_>$ , where the subscript `$>$' denotes
the positive (polynomial in $x$) part of the Laurent expansion around
infinity.  Note that $n$ is of order $1$ and $U_+$ is known (to be
computed below) such that the task of finding $P_{n-1}(x;u)$ is
relatively simple.

Eq.  (\ref{F0Solution}) represents a solution for $F^{(0)}$ given the
decomposition (\ref{Decomposition}).  Fortunately, the functions
$U_\pm$ may be computed explicitly.  Assume that the phase of the
complex function $1-\alpha \CQ(x)$ winds $n_a$ times as $x$ moves
around the $a$-th arc once (where we use the conventions of positive
winding for counterclockwise rotation).  Namely, we assume that the
rational function $1-\alpha \CQ(x)$ has $n_a$ more zeros than poles a
microscopic distance around the $a$-th arc.  Below we shall call $n_a$
below simply 'the winding number'.  Let $n=\sum_an_a$.  For each $a$
we find a rational function $R_a(x)$ such that [$1-\alpha
\CQ_\ii(x)]R(x)$ has winding number $0$ around all arcs, where
$R(x)=\prod_a R_a(x)$.  We choose the functions $R_a(x)$\ as follows.
If $n_a<0$, we take $R_a(x) =\prod_{i=1}^{|n_a|} (x-u_{j^{(a)}_i}), $
where $u_{j^{(a)}_i}$, $i=1, \dots, |n_a|$, are arbitrary roots
belonging to the $a$-th arc.  The final result at given order does not
depend on this choice to the corresponding order.  If $n_a>0$, we
choose $R_a(x) =\prod^{n_a}_{i=1}\frac{1}{x-\alpha_{j^{(a)}_i}}, $
where $\alpha_{j_i}^{(a)}$ are a set of $n_a$ roots of $1-\alpha
\CQ(x)$ around the $a$-th arc.  The functions $U_+(x)$ and $U_-(x)$
are then given by
\begin{align}   
U_+ (x)&=\frac{\exp \left\{\CP_\uu\left[\log{\left(1-\alpha
  \CQ(x)\right) R(x)}\right]\right\}}{R(x)} ,
\label{UmDef} \\
U_-(x) &=\frac{1-\alpha \CQ (x)}{U_+(x)}.
\label{UpDef}
\end{align}

\subsection{Semi-classical  expansion of the fast piece}  
 
To obtain the semiclasical expansion of the first term in
(\ref{GeneralMain}), we need to be able to expand the expression
$\oint e^{-\Phi}\log\left(1-\alpha \ID_\ii\right)e^{\Phi}.$ This is
possible only if the contour of integration $\CC $ is far from both
$\uu$ and $\zz$.  At leading order
\begin{align} 
\oint\limits_\CC {dx\over 2\pi }\ e^{-\Phi}\log\left(1-\alpha \ID_\ii
\right) e^{\Phi} =\oint\limits_\CC {dx\over 2\pi }\ \log\left(1-\alpha
\CQ(x) \right) +\CO(1/L).
  \label{simplestLog}
\end{align}
The $O(L^0)$ correction is an integral of pure derivative and
vanishes.  Note that the integrand on the right hand side has
logarithmic branch cuts emanating from the arcs formed by the points
from the set $\uu$ whenever the winding number $n_a$ of the function
$1-\alpha \CQ$ around the $a$-th arc is non-zero.

By construction, the function $\CQ(x)$ has no winding numbers, but
only cuts along the arcs.  The cuts appear as the result of merging of
the poles and the zeros of $\Qii(x)$ in the limit $\ii\sim 1/L \to 0$.
 When $\a$ is small, the function $1-\alpha \CQ_\ii$ has no zeros near the 
 cuts   and the contour of integration may be drawn to simply encircle the  cuts.

 As $\alpha$ increases, a number of zeros of $1-\alpha \CQ_\ii$ on the
 second sheet can move through the cuts to the first sheet.  The
 contour of integration in (\ref{simplestLog}) should be drawn to
 surround those zeros.  This presents no problem as moving the contour
 away from the arc is consistent with the expansion in
 (\ref{simplestLog}).  If, on the other hand, an extra zero (one that
 was not there at small $\alpha$) of $1-\alpha \CQ_\ii$ approaches the
 $a$-th arc, as $\alpha$ increases, the contour of integration must be
 drawn between that zero and the arc.  Eventually, that zero may
 approach the arc up to a microscopic distance, and the approximation
 leading to (\ref{simplestLog}) will be invalidated.  To deal with
 this scenario one must separate out he roots around the zero and
 compute their contribution to the fast piece by performing the sum in
 (\ref{FLi2})\ for those roots more directly.  We do not show how this
 is done explicitly in this paper, rather it will be the subject of
 future work.

\subsection{The leading order result}

Only the fast piece contributes to the leading order ($\sim L$) of the 
semiclassical expansion of
  $\log\mathscr{A}_{\uu,\zz}^{[\ii ,\kappa]}$, since the slow 
  piece can be easily seen to be
  of order $L^0$:
  \begin{align}
\log\mathscr{A}_{\uu,\zz}^{[\ii ,\kappa]} =\int_0^\kappa 
\frac{d\alpha}{\alpha} \oint \limits_\CC {dx\over 2\pi i}\
\log\left[1-\alpha \CQ(x)\right],   
\end{align}  
where the  branch cut of the logarithmic function is to be taken 
according to the prescription in the previous subsection. 
Sometimes   the contour should be deformed so that part of it passes 
in the second  sheet,  as  explained  in \cite{SL}.
In any case, the integral over $\alpha$ can be taken and  the final result is
\begin{align}
\log\mathscr{A}_{\uu,\zz}^{[\ii ,\kappa]} =- {1\over\ii}
 \oint\limits_{\CC} {dx\over 2\pi i} 
\
\Li\left(\kappa Q(x)\right).
\end{align}

\subsection{Subleading order for  zero winding numbers }

In this section we will assume that $n_a=0$ to avoid the complications
that arise in the case of non-vanishing winding numbers.  With this
assumption we will write a compact expression for the leading and the
subleading orders.  Combining (\ref{F0Solution}) and (\ref{TrFz}),
with $P_{n-1}(u)=0$ (which is appropriate for $n_a=0$), we obtain at
leading order
\begin{align}
&\sum_j \<j|\mathcal{F}^{\slow }|j\>=-
 \oint\limits_\CC \frac{dx}{2\pi i}\oint\limits_\CC \frac{du}{2\pi i} 
  \frac{1}{U_+(x)U_-(u )(x-u )^2}= \nonumber\\
&=-\oint\limits_\CC \frac{du}{2 \pi i } \ \frac{\p_u
U_+(u)}{U^2_+(u)U_-(u)} =-\oint\limits_\CC \frac{du}{2 \pi i } \
\frac{\p_u\log\left[U_+(u)\right]}{\left(1-\alpha Q(u)\right) } .
\end{align}
Using the explicit definition of $U_+, $ Eq.  \re{UmDef}, where we
take $R(u)=1$, this can be further written as
\begin{align}
&\sum_j\<j|\mathcal{F}^{\slow}|j\> = \oiint\frac{dxdu}{(2 \pi i)^2} \
\frac{1}{1-\alpha \CQ(x)} \frac{1}{(x-u)^2} \log\left( 1-\alpha \CQ(u)
\right) = \nonumber\\
&=- \frac{\alpha}{2}\partial_\alpha \oiint  \frac{dxdu}{(2 \pi i)^2} \ 
\log\left( 1-\alpha \CQ(x)  \right) \frac{1}{(x-u)^2} \log\left( 1-\alpha \CQ(u) \right). 
\end{align}
Adding the contributions from the fast and the slow pieces, we arrive
at the following approximation for $\log\mathscr{A}_{\uu,\zz}^{[\ii
,\kappa]}$, correct to order $O(1):$
\begin{align}
 \la{leadsubRH} \log\mathscr{A}_{\uu,\zz}^{[\ii ,\kappa]} &=- {1\over
 \ii} \oint\limits_\CC {dx\over 2\pi i} e^{-\Phi(x)} \,
 \Li\left(\kappa \ID_\ii\right) e^{\Phi(x)} + \nonumber\\
 &+\frac{1}{2}  \oiint\limits_{\CC\times \CC}\frac{dx du}{(2 \pi i )^2}  \log\left[ 1-\kappa \CQ(x) \right]
 \frac{1}{(x-u)^2} \log\left[ 1-\kappa \CQ(u) \right].
\end{align}
The first term on the right hand side is easily expandable in $1/ L$
as the contour of integration can be taken to be well away from the
arcs formed by the points of the set $\uu$.  The second term is an
$\CO(1)$ correction.  Higher order correction are straightforward to
compute by incorporating the contribution of $F^{(n)}$ for $n>0$.  As
mentioned above, the case when some winding numbers are non-zero
implies more complicated expressions, which we do not develop here.

\section{Effective field theory for the semiclassical limit}
\la{FieldTheory}

%We  can  rewrite the sum \re{ExpAb}  in terms of multiple
%contour integrals 
%%
%  \begin{eqnarray}
%%  \la{expFD}
%  \begin{aligned}
%  \caA_{\uu, \zz }^{[\ii, \k]}&= \sum_{n=0}^N {(-\k)^n\over n!}
%  \prod_{j=1}^n \oint_{\CC_\uu} {dx_j \over 2\pi i} \ \ {\Qii (x_j)
%  \over \ii}\ \ \prod_{k>j}^n { (x_{j}-x_k)^2 \over (x_{j}-x_k)^2- \ii
%  ^2 } \, .
%    \end{aligned}
%  \end{eqnarray}
%  %
% In this formulation, the semi-classical limit of the $\caA$-functional
%resembles the so-called Nekrasov-Shatashvili limit of the instanton
%partition functions of deformed $\CN=2$ supersymmetric gauge theories
%\cite{Nekrasov:2009aa}, for which well developed techniques exist.
%The representation of the $\caA$-functional as a multiple contour
%integral can be interpreted as a non-ideal one-dimensional gas of
%particles (dipoles) confined on the contour $\CC$.  The free energy of
%this gas is obtained by iterated Mayer expansion.  The procedure,
%outlined in \cite{Nekrasov:2009aa}, was recently worked out in great
%detail in \cite{Mayer-MY, JEB-Mayer}.

%Instead of delving into the diagrammatics of the cluster expansion, we
%will explain the procedure in field-theoretical terms.  
In this section we 
reformulate the $\caA$-functional in terms of a chiral fermion
or, after bosonization, in terms of a chiral boson with exponential
interaction.  The interaction is weak at large distances but becomes 
singular at distances of order $\ii$, where the two-point function develops poles. 
 Our goal is to formulate an effective field theory for the limit $\ii\to 0$.
 For that we split the theory into a fast and a
slow component and integrate with respect to the fast component.
%, with
%the result that at large distances the effective action contains also
%exponential fields of higher dipole charge, which emerge in the
%operator product expansion of the original exponential fields.  
%The
%semiclassical limit is described by an effective infrared theory, in
%which the exponential fuelds creating the fundamental and the
%composite particles are treated on equal footing.  The classical
%action for the effective theory is analogous to the Yang-Yang
%potential in the sense of ref.  \cite{Nekrasov:2009aa}.

  \subsection{Free fermions}
  
This determinant \re{detAK} is a particular case of the
$\tau$-functions considered in section 9 of \cite{JimboMiwa-tau} and
can be expressed as a Fock-space expectation value for a Neveu-Schwarz
chiral fermion living in the rapidity complex plane and having mode
expansion
\be \la{pzpoa} \psi (u)= \sum_{r\in \IZ+ {1\over 2}}\psi_{r}\ u^{-r-
{1\over 2}}, \ \ \ \psi^* (u)= \sum_{r\in \IZ+ {1\over 2}} \psi^*_{
r}\ u^{r- {1\over 2}} .  \ee
The fermion modes are assumed to satisfy the anticommutation relations
\be\la{cpmtoa} [\psi_{r},\psi^*_{s}]_+=\delta_{rs}\, , \ee
and the left/right vacuum states are defined by
\be\la{mnfio2a} \bra \psi_{-r}= \bra \psi^*_{r} = 0\ \ \text{and}\ \ \
\psi_{r}\, \ket = \psi^*_{-r} \ket = 0,\ \ \ \text{for} \ r> 0 .  \ee
The operator $ \psi^*_r$ creates a particle (or annihilates a hole)
with mode number $r$ and the operator $\psi_r$ annihilates a particle
(or creates a hole) with mode number $r$.  The particles carry charge
1, while the holes carry charge $-1$.  The charge zero vacuum states
are obtained by filling the Dirac see up to level zero.

Any correlation function of the operators \re{pzpoa} is a determinant
of two-point correlators
\be \la{opepsia} \bra \psi(u) \psi^*(v)\ket = \bra \psi^*(u)
\psi(v)\ket = {1\over u-v}\, .  \ee
The expectation value of several pairs of fermions is given by the
determinant of the two-point functions.  Obviously the determinant
\re{detAK} is equal to the expectation value
 \be \la{fermionrepD} \Det(1-\k K) = \bra \exp \(\k \ii \sum_{j=1}^N
 E_j \, \psi^*(u_j ) \psi(u_j+ \ii)\)\ket .  \ee
The discrete sum of fermion bilinears in the exponent on the rhs of
\re{fermionrepD} can be written, with the help of \re{EjQ}, as an
integral along the contour $\CC_\uu$ which encircles the points
$u_1,\dots, u_N$, and the Fock space representation \re{fermionrepD}
takes the form
 \be \la{fermionrepC} \caA_{\uu,\zz}^{[\ii, \k]} = \bra \exp \({\k}
 \oint_{\CC_\uu} {d x\over 2\pi i} \Qii (x) \, \psi^* (x )
 \psi(x+\ii)\)\ket, \ee
where the weight function $\CQ_\ii(x)$ is defined by Eq. \re{defQii}.

\subsection{Bosonic  field with exponential interaction}

Alternatively, one can express the $\caA$-function in term of a chiral
boson $\phi(x)$ with two-point function
\be
\bra \phi(x) \phi(y)\ket = \log(x-y).
\ee
After bosonization $\psi(x) \to e^{\phi(x)}$ and $\psi^*(x) \to
e^{-\phi(x)}$, where we assumed that the exponents of the gaussian
field are normally ordered, the fermion bilinear $\psi^*(x) \psi(x+
\ii)$ becomes, up to a numerical factor, a chiral 
exponential field\footnote{Our convention is  that the exponential is normally ordered,
$e^{\phi(u)-\phi(v)} \equiv\  :e^{\phi(u)-\phi(v)}:$\, , \ so that $\< 0|e^{\phi(u)-\phi(v)} |0\>=1$.}
\be
\la{defVeps}
\CV_\ii(x) \equiv e^{\phi(x+\ii) - \phi(x)}.
\ee
The numerical factor is determined by the OPE
\be 
\la{opephi}
e^{-\phi(x)} \, e^{\phi(u)} ={1\over x-u} e^{\phi(u)-\phi(x)}
\qquad (u= x+\ii)\, ,
\ee
so that the  fermion bilinear  bosonizes as
  \be \la{defAvert} \psi^*(x) \psi(x+ \ii)\quad \ \to\ \quad
  e^{-\phi(x)}e^{\phi(x+\ii) } = - {1\over \ii}\, \CV_\ii(x) .  \ee
The resulting bosonic field theory is that of a two-dimensional
gaussian field $\phi(x,\bar x)$  perturbed by a chiral interaction term $\k\Qii (x)\CV_\ii(x)$.  
%The interaction term, which is proportional to  the  
%exponential field \re{defVeps}, 
%creates a pair of Coulomb charges with opposite signs spaced at distance
%$\ii$.  
 
 Expanding the exponential in series and using that 
the $n$-point correlator of  the exponential field is a product of all two-point 
correlators  
 \be
 \la{pairwiseV}
  \bra \Vii(x) \Vii (y)\ket = { (x-y)^2 \over (x-y+\ii)(x-y-\ii)} \, ,
 \ee
 we  obtain  that the expectation value \re{fermionrepC} is given by the
grand-canonical Coulomb-gas partition function
  \begin{eqnarray}
 \la{expFD}
  \begin{aligned}
  \caA_{\uu, \zz }^{[\ii, \k]}&= \sum_{n=0}^N {(-\k)^n\over n!}
  \prod_{j=1}^n \oint_{\CC_\uu} {dx_j \over 2\pi i} \ \ {\Qii (x_j)
  \over \ii}\ \ \prod_{k>j}^n { (x_{j}-x_k)^2 \over (x_{j}-x_k)^2- \ii
  ^2 } \, .
    \end{aligned}
  \end{eqnarray}

 In the contour integral representations \re{fermionrepC}  and  \re{expFD}
 the integration contour
$\CC_\uu$ is drawn close to the poles of $\Qii $.  
We would like to
deform the contours away  from these poles, 
where $\Qii$ can be considered as a sufficiently smooth.  
%If we expand
%the contours in the series we will gradually encounter new poles of
%the integrand at $x_j-x_k = \pm \ii$.  These new poles will add to the
%poles of $\Qii$ and will arrange in strings $\{ u_j, u_j+\ii, \dots,
%u_j +n\ii\}$.  A general configuration contains several such strings.
%The partition function can be evaluated by summing up the residues of
%all such (multi-)string configurations.  An alternative to that is the following.
 In  the
$n$-th term of the series we can  deform sequentially the integration contours  away from the set $\uu$, so that the $n$ contours form a nested
configuration separating the $\uu$-poles and the $\zz$-poles of the
function $\Qii$.  If  the subsequent contours are spaced by $\ii$, then 
the poles at $x_j-x_k=\pm \ii$ of the integrand do not contribute.

   \subsection{Integrating out the fast modes}

%After that we further deform the $n$ contours to the
%same contour $\CC$, taking into account the residues at the poles.
% This is equivalent to the cluster (Mayer) expansion.  The
%argument that follows is a field-theory version of the cluster
%expansion.  
%  
% 
%The interaction describes a Coulomb gas of `fundamental' particles
%(dipoles) 
%with zero electric charge but non-vanishing dipole and higher charges.
% The  dipoles
%interact with the external potential $\Phi(x)$ and pairwise among
%themselves.  The pairwise interaction is  given by  the two-point correlator 
%\re{pairwiseV}.
%%%
%% \be \bra \Vii(x) \Vii (y)\ket = { (x-y)^2 \over (x-y+\ii)(x-y-\ii)} .
%% \ee
%%%
%Subtracting the product of the one-point functions $\bra \Vii \ket=
%1$, one obtains for the connected correlator of two dipoles
%%
%\be \la{vavA} \langle \!\langle \Vii (x)\Vii (y)\rangle\!\rangle =
%{\ii^2\over (x-y)^2 - \ii^2}.  \ee
%%
%The interaction between two dipoles depends both on the distance and
%on the direction.  If $\ii=|\ii|i$, then the force between two dipoles
%is repulsive if they are spaced horizontally and attractive if they
%are spaced vertically.
% 
% 
% 
%As the interaction rapidly decreases at large distances, one can
%compute the thermodynamics of the dipole gas by performing iterated
%Mayer expansion as in \cite{Mayer-MY }.  In field theory terms, this
%amounts to split the field into a fast and a slow components, $\phi(x)

The multiple contour integral can be evaluated in the semiclassical limit by 
splitting the integrand into slow and fast parts.  
%We will not go into the details 
%of the method, which can be found {\it e.g.}  in \cite{Mayer-MY}.
%After evaluating exactly the fast part of the integral, the Coulomb gas partition
% function is expressed as  a multiple integral along 
%{ \it single} contour $\CC$, but  with modified integrand.
 We thus introduce an intermediate scale $\L$ such that 
\be
|\ii|\ll\L\ll
N|\ii|\, 
\ee
and split the bosonic field into a fast and a slow components, 
\be
\phi= \phi_{\slow} + \phi_{\fast}.
\ee
Up to exponential terms the two-point function of the bosonic field is approximated
at small  distances  by 
that of the fast component and at  large distances   
by that of the slow  component.
Below we will perform explicitly the integration with respect to  $\phi_\fast$
to obtain an effective interaction for $\phi_\slow$.
Since the two-point function of the exponential fields with 
$\phi$ replaced by $\phi_\slow$ does not contain poles, the  nested
contours spaced by $\ii$ can be replaced by  a {\it single } contour $\CC$
placed sufficiently far from the sets $\zz$ and $\uu$ where the integrand 
has poles.\footnote{ The
splitting into a fast and a slow components can be done explicitly if
the contour $\CC$ can be placed along the real axis.  Introduce a
cutoff $\L$ such that $|\ii|\ll \L \ll N|\ii|$.  Then the bosonic
field has a continuum of Fourier modes $\a_E$ and the slow and fast
parts can be defined as $\phi_{\slow} (x)= \int_{|E|<\L} dE\,  \a_E \
e^{i E x}, \phi_{\fast}(x) = \int_{ |E|>\L} dE \, \a_E\ e^{iE x}$.  The
propagators of the slow and the fast components are $ \bra \p
\phi_{\slow} (x) , \phi_{\slow} (y)\ket = (1- e^{i \L (x-y)})/(x-y),\
\bra\p \phi_{\fast} (x) , \phi_{\fast}(y)\ket = e^{i \L (x-y)}/(x-y).$
The propagator for the slow component contains a strongly oscillating
term whose role is to kill the pole at $x=y$ and which can be
neglected far from the diagonal, while the numerator in the propagator
of the fast component can be replaced by 1 at small distances.  The
effects of the cutoff are thus exponentially small and do not
influence the perturbative quasiclassical expansion.  }
% and  split the bosonic  field into a fast and a slow components, 
%%
%\be
%\phi(x)= \phi_{\slow} + \phi_{\fast},
%\ee
%%
%having wave lengths respectively larger and smaller than the cutoff 
%$\Lambda$.% At small distances (with respect to $\L$) the dynamics is
%determined only by the pairwise interactions and the slowly varying
%external potential can be treated as an entire function.  Then the
%theory becomes solvable.  Integrating with respect to the fast degrees
%of freedom, one obtains an effective theory for the slow degrees of
%freedom which are relevant for the infrared behaviour. We will do
%that at a reasonably intuitive level; a rigorous procedure can be
%developed in close analogy with the analysis of \cite{Nekrasov:2009aa,
%Mayer-MY, JEB-Mayer}.
%
% We split the integration domain into cells of size $\Lambda$ 
% and perform the integration in each of the cells, neglecting 
% terms exponentially small in $\Lambda$.  The integration  
% within a cell can be done by residues.
The effective interaction for the slow component is 
of the form 
%%
%\be
%S_\eff[\phi_\slow]=\sum_{n\ge 1}
%\oint_\CC {dx\over 2\pi i} \  \Xi_n(x)\, ,
%\ee
%%
%
\be
S_\eff[\phi_\slow]=\sum_{n\ge 1}
\oint_\CC {dx\over 2\pi i} \ V_\eff ^{(n)}(x)\, ,
\ee
where $n$-th term  is  the  contribution of the connected
$n$-point function of the exponential field $\CV_\ii(x)$ with respect to the fast component,
 \be
 \la{Veffslow}
 V_\eff ^{(n)}(x)=
{\({-\k/ \ii} \)^n\over n!}  \oint {dx_1\over 2\pi i} \dots   {dx_n\over 2\pi i} 
\ \d( x-x_1)
\Big\<\!\!\Big\<\prod_{j=1}^n  \  \CQ_\ii(x_j) \, \CV_\ii(x_j)
\Big\>\!\!\Big\>_{\!\! \fast} .\ee
 %
% %
 \be
 \la{Veffslow}
 \Xi_n(x)=
{\({-1/ i\ii} \)^n\over n!}  \oint {dx_1\over 2\pi i} \dots   {dx_n\over 2\pi i} 
\ \d( x-x_1)
\Big\<\!\!\Big\<\prod_{j=1}^n  \  \CQ(x_j) \, \CV(x_j)
\Big\>\!\!\Big\>_{\!\! \fast} .\ee
% %

 To compute the connected correlation function $\<\!\<\ \>\!\>_{\fast}$
we represent  the product of $n$ exponential fields in the form
\be
\la{OPE}
\Vii(x_1)\dots \Vii(x_n)=
\prod _{j<k} {(x_j-x_k)^2
    \over (x_j-x_k)^2- \ii^2}\
:\Vii(x_1)\dots \Vii(x_n):\, ,
\ee
 where $:\ :$ signifies normal product. By definition the normal product of exponential fields has vacuum expectation value 1.  Assuming  that all distances  $|x_j-x_k|$ are small compared to the scale $ \L$, we can interpret the operator product expansion \re{OPE} as 
    \be \la{OPEA} 
    \big\< \Vii(x_1)\dots \Vii(x_n)\big\>_{\!\fast}\approx  \prod _{j<k} {(x_j-x_k)^2
    \over (x_j-x_k)^2- \ii^2}\
    : \Vii^\slow(x_1)\dots \Vii(x_n)^\slow :\, .\ee
    \be \la{OPEA} 
    \big\< \CV(x_1)\dots \CV(x_n)\big\>_{\!\fast}\approx  \prod _{j<k} {x_{jk}^2
    \over x_{jk}^2+ \ii^2}\
    : \CV^\slow(x_1)\dots \CV(x_n)^\slow :\, .\ee
To extract the connected component of the $n$-point function 
we apply  the Cauchy identity
\re{cauchy} and represent  the Cauchy determinant as a sum over 
permutations.
 The result is the sum of the (identical) contributions of the 
  $(n-1)!$ permutations representing maximal cycles
of length $n$.\footnote{This is basically the calculation done in   ref. \cite{G.Moore:2000aa}.
The difference is in the extra factor and in our convention to choose $x=x_1$ 
as collective coordinate, while in \cite{G.Moore:2000aa} $x= (x_1+\dots+x_n)/n$.
  Note that the contribution of the 
permutations with more than one cycle vanishes
automatically.}     
%We integrate \re{OPEA}
%along a  segment of the contour $\CC$ of size $\Lambda$ and 
%containing the point $x$.
%Since we want to evaluate the effect of the short-distance interaction
%due to the poles, we can assume that the rest of the integrand is
%analytic everywhere.  Then the integration can be performed by
%residues using the Cauchy identity.  The easiest way to compute the
%integral is to fix $x_1=x$ and integrate with respect to $x_2, \dots,
%x_n$.  We expand the numerical factor in \re{OPEA} as a sum over
%permutations.    For the
%rest of the permutations the contour integral vanishes.  
We find
($x_{jk} \equiv x_j-x_k$)
\be
 V_\eff ^{(n)}(x_1)&=&
%  \oint { \Vii(x_1)\dots \Vii(x_n)\over (-\ii)^n \ n!} \prod_{k=2}^n
%{dx_k\over 2\pi i}\ 
%&\sim & 
{ (n-1)!\over n!} \oint { \prod_{k=1}^n
\CQ_\ii(x_k) \, \CV(x_k)^\slow  \
{dx_2\over 2\pi i}\, \dots {dx_n\over 2\pi i}\, 
\over (\ii-x_{12} )
\dots(\ii-x_{n-1,n})(\ii- x_{n,1})} 
\no
\\
\no
\\
&=&-
{Q_{n\ii}(x_1) \over n^2\ii} \   \CV_{n\ii}(x)\, ,
\ee
with
\be \CQ_{n\ii}(x) &\equiv& 
Q_{\ii}(x) Q_\ii(x+\ii) \dots Q_\ii(x+ n\ii) \ = e^{-\Phi(x)+\Phi(x+n\ii)} \, ,
\\
 \CV_{n\ii}(x) &\equiv& 
 : \CV_{\ii}(x)  \CV_{\ii}(x+\ii)\dots   \CV_{\ii}(x+n\ii):\ =
  e^ {-\phi(x) + \phi(x+ n\ii)}. 
\la{compositep} \ee
(Here and below  $\phi$ denotes  $\phi_\slow$.)
%These composite particles are electrically neutral, have dipole
%charges $n$ times the dipole charge of the fundamental particles and
%interact with the external potential through the weight factors
%$\CQ_{n\ii}(x)/n^2\ii$, with
%%
%\be \CQ_{n\ii}(x) = e^{-\Phi(x)+\Phi(x+n\ii)} \, .  \ee
%%
 The resulting expression for the  $\caA$-functional  in terms of 
 the effective  infrared  theory is
 \be \la{effectiveFT} \mathscr{A}_{\uu,\zz}^{[\ii ,\kappa]} = \bra
 \exp \(-{1\over\ii} \sum_{n=1}^ {\L/\ii} {\k^n\over n^2}
 \oint_{\CC_\uu} {d x\over 2\pi i} \CQ_{n\ii} (x) \,
 \CV_{n\ii}(x)\)\ket + \text{non-perturbative}. \ee

 By construction the spacing $n\ii$ should be smaller than the scale $\L$,
 but if the sum  over $n$ in
the exponent is extended  to infinity,  this will introduce exponentially small
terms and will not change the $1/L$ expansion.
Introducing the shift operator $\ID_\ii= e^{\ii \p_x}$,
the series in the exponent can be formally summed up   as
\be
\la{dilogoper}
\begin{aligned}
\mathscr{A}_{\uu,\zz}^{[\ii ,\kappa]} 
%&= \bra \exp\(-{1\over \ii}
%\oint_{\CC} {dx\over 2\pi i} \ \sum_{n=1}^\infty {\k^n\over n^2}
%:e^{-\Phi(x) - \phi(x)} \ e^{\Phi (x+n \ii ) +\phi(x+n \ii)}: \) \ket
%\\
&= \bra \exp\(- {1\over \ii} \oint\limits _{\CC} {dx\over 2\pi i} \ :e^{
-\Phi(x)-\phi(x) } \ \Li(\k \, \ID_\ii)\ e^{ \Phi(x) +\phi(x)}: \)
\ket \, 
+ \text{non-perturbative}.
\end{aligned}
\ee
 Another way to write this expression, without using the normal product
 and  redefining $\phi+\Phi\to \phi$, is
\be
\la{nonormal}
\begin{aligned}
\mathscr{A}_{\uu,\zz}^{[\ii ,\kappa]} 
%&= \bra \exp\( 
%\oint_{\CC} {dx\over 2\pi i} \ \sum_{n=1}^\infty {\k^n\over n}
%e^{-\Phi(x) - \phi(x)} \ e^{\Phi (x+n \ii ) +\phi(x+n \ii)} \) \ket
%\\
&= \bra \exp\Big(-   \oint\limits _{\CC} {dx\over 2\pi i} \, e^{
-\phi(x) } \ \log(1-\k\,  \ID_\ii)\ e^{ \phi(x)}: \Big)
\ket \, 
+ \text{non-perturbative},
\\
& \bra \phi(x)\ket   =\Phi(x), \ \ 
\bra \phi(x)\phi(y) \ket = \log (x-y).
\end{aligned}
\ee

\subsection{One-dimensional effective theory in the semiclassical
limit}

 In the semiclassical
limit
\be \la{semicllim} \hbar \equiv 1/ L\to 0 , \qquad \ell \equiv
L\ii\sim 1, \qquad \a = N/L\sim 1 \ee
the classical field $\Phi$  grows as $1/\hbar$, but
\be \la{diffPhi} \Phi(x+n\ii)-\Phi(x) = n \ii\, \p\Phi(x) + \dots \ee
remains finite, as well as   the range of
integration and size of the contour $\CC$.      
Furthermore, the distribution of the roots $u_j$ is assumed to be of
the form of the finite zone solutions of the Bethe equations
\cite{Kazakov:2004qf}, which are described by hyperlliptic curves.
The roots $\uu$ condense into one or several arcs, which become the
cuts of the meromorphic function \be \la{defp} \p\Phi (x) = \sum
_{j=1}^N {1\over x-u_j}-\sum_{l=1}^L {1\over x- z_l} .  \ee
We assume that the inhomogeneities $\zz$ are centered around the
origin of the rapidity plane, but we do non make any other assumptions
about them.

We will concentrate on the leading term (of order $1/\ii$) and the
subleading term (of order 1), and will ignore the corrections that
vanish in the limit $\ii\to 0$.  Then, using the approximation
\re{diffPhi}, we expand the exponent in \re{dilogoper} as
\be
\la{approxFT}
\begin{aligned}
\mathscr{A}_{\uu,\zz}^{[\ii ,\kappa]} & = \bra \exp \oint_{\CC}
{dx\over 2\pi i}\( - {1\over \ii}: \Li (\k \, \CQ\, e^{ -\ii\vp}) : - :
\log(1- \CQ \, e^{-\ii \vp})\p\vp : + \dots \) \ket \, ,
\end{aligned}
\ee
where we introduced the derivative field
\be \vp(x) = - \p\phi(x)
 \ee
and used the notation \re{QcalisQ}.
We can retain only the first term on the rhs of \re{approxFT}, since
the second term is a full derivative and can be neglected.
 
Now we can pass from Fock-space to path-integral formalism.  For that
we express the expectation value \re{approxFT} as a path integral for
the $(0+1)$-dimensional field $\vp(x)$ defined on the contour $\CC$
and having two-point function
\be \la{tpf} G(x,u)\equiv \< \vp(x)\vp(u)\> = \ {1\over (x-u)^2}\, .
\ee
Introducing a second  field $\rho $ linearly coupled to $\phi$
we write the $\caA$-functional as a path integral
\be \mathscr{A}_{\uu,\zz}^{[\ii ,\kappa]} &=& \int [D\vp\, D\rho]\
e^{- \CY[\vp, \rho]}\, ,
\la{pathint}
 \ee
with the action functional given by
   \be \CY[\vp, \rho]= \oint\limits_\CC {dx \over 2\pi i}\( {1\over
   \ii}\Li(\k \CQ(x) e^{-\ii\vp(x)})+\vp(x)\rho(x)\) + \hf
   \oint\limits_{ \CC\times\CC} {dx\, du \over (2\pi i)^2}\ \rho(x)
   G(x, u )\rho(u).  \la{defYY} \ee
The double integral in the second term can be understood as a
principal value.  Indeed, the contribution $\rho\rho'$ of the pole at
$x=u$ is pure derivative and vanishes after being contour-integrated.

In the approximation we are looking for, the $\caA$-functional is
given by the saddle-point action
 \be \log \caA^{[\ii, \k]} _{\uu, \zz}&= {\CY_{c}} + \CO(\ii), \qquad
 \CY_{c}= \CY[\vp_c, \rho_c], \ee
where the saddle point $\vp _c$ is given by a couple of TBA-like
equations
\begin{eqnarray}
\la{TBAlike} \vp_c(x) = - \!\!\!\!\!\!  \int\limits_\CC {dy\over 2\pi
i} G(x-y) \rho_c(y), \quad \rho_c(x) = - \log\( 1-\k \CQ(x)
e^{-\ii\vp_c(x)}\).
\end{eqnarray}
After solving for $\rho_c$, one obtains a non-linear integral
equation\footnote{Such type of integral equations first appeared as
alternative formulation of the Thermodinamic Bethe Ansatz without
strings \cite{Destri:1994bv, 0305-4470-24-13-025}, and most recently
in supersymmetric gauge theories \cite{Feverati:2006tg,
Nekrasov:2009aa}. If the space-time variable $x$
 scales as $\ii^0$,  there is no need to solve the non-linear integral equation,
because only the leading order in $\ii$ matters.
We don't exclude that the above analysis can be carried on for 
weaker assumptions about the distribution of the roots $\uu$, such that 
 $x$ scales as $\ii^1$, in which case the non-linear integral equation does not contain a small parameter.}
 for the classical field $\vp_c$:
\begin{eqnarray}
\la{NLIE} \vp_c(x) =- \!\!\!\!\!\!  \int\limits_{\CC}{dy\over 2\pi i}
G(x-y) \log\( 1-\k \CQ(y) e^{-\ii \vp_c(y)}\) .
\end{eqnarray}
 Expanding
\begin{eqnarray}
\la{YYcrit} \CY _c &=& \oint\limits_{\CC}{dx\over 2\pi i} \
 \[ -{1\over\ii}\Li(\k \CQ(x) e^{-\ii \vp_c(x)})
 -\hf \vp_c(x) \log\( 1-\k \CQ(x) e^{-\ii \vp_c(x)}\) 
\]
\end{eqnarray}
up to $O(\ii)$, we obtain an explicit expression for the leading and
the subleading terms:
 \be \la{leadsubleadQFT}
 \begin{aligned}
   \hskip -0.3cm \log \caA^{[\ii, \k]} _{\uu, \zz} &= - {1\over\ii}
   \oint \limits_{\CC} {dx\over 2\pi i} \ \Li[\k \CQ(x)] + \hf
   \oint\limits_{ \CC\times\CC} {dx\, du \over (2\pi i)^2}\ { \log\[
   1-\k \CQ(x) \] \ \log\[ 1-\k \CQ(u) \]\over (x-u)^2} \\
&+\CO(\ii),
 \la{Aleading}
\end{aligned}
\ee
where the double integral is understood as a principal value. 

The expression \re{leadsubleadQFT} obtained by the field-theory method
is identical to the result obtained by solving the Riemann-Hilbert
problem, Eq.  \re{leadsubRH}.  Taking $\ii=i$ and $\CQ=
\exp(ip_\uu+ip_\vv)$, we obtain the expression for the leading and the
sub-leading terms of the inner product, Eqs.  \re{CFzero}--\re{CFone}.
Here we neglected the trivial factors in the expression \re{caAuvuz} of
the inner product through the $\caA$-functional.

The choice of the contour $\CC$ is a subtle issue and depends on the
analytic properties of the function $\CQ(x)$, as discussed above in
Section \ref{RHApproach}.  In any particular case one can first find
explicitly the function $\CQ(x)$ in the limit of a small filling
fractions ($\a \equiv N_a/L\ll1$, where $N_a$ is the number of roots
that form the $a$-th arc), then place the contour $\CC$ so that it
does not cross any cuts of $\Li(\CQ(x))$.  If the fillings are not too
large, this choice of the contour will remain valid also for
$N_a/L\sim 1$.  However, it is possible that at some critical filling
that one of the the zeros of $1-\CQ$ approaches the $a$-th arc.  Such
a situation has been analysed in \cite{Bargheer:2008kj}.  If this is
the case, the contour of integration should be deformed to avoid the
logarithmic cut starting with this zero, possibly passing to the
second sheet.
  
  \subsection{Relation to the Mayer expansion  of non-ideal  gas}

  The semi-classical limit of the $\caA$-functional
resembles the so-called Nekrasov-Shatashvili limit of  instanton
partition functions of deformed $\CN=2$ supersymmetric gauge theories
\cite{Nekrasov:2009aa}.  The methods developed to study this limit,
outlined in \cite{Nekrasov:2009aa} and recently worked out in great
detail in \cite{Mayer-MY, JEB-Mayer}, are based on the iterated Mayer expansion
for a non-ideal gas of particles confined along a contour $\CC$.
Below we are going to explain the connection between our approach and the 
Mayer expansion.

The exponential field \re{defVeps} creates  a
 pair of Coulomb charges with opposite signs spaced at distance
$\ii$.  One can think of such a pair as a `fundamental' particle
with zero electric charge but non-vanishing dipole and higher charges.
 The sum \re{expFD} is the grand partition function    of such  `fundamental'  dipoles
 confined on the contour $\CC_\uu$, or equivalently, on a sequence of nested contours
  surrounding the set $\uu$.

The  fundamental dipoles
interact with the external potential $\Phi(x)$ and pairwise among
themselves.  The pairwise interaction is  determined  by  the two-point correlator 
\re{pairwiseV}.
%%
% \be \bra \Vii(x) \Vii (y)\ket = { (x-y)^2 \over (x-y+\ii)(x-y-\ii)} .
% \ee
%%
Subtracting the product of the one-point functions $\bra \Vii \ket=
1$, one obtains for the connected correlator of two dipoles
\be \la{vavA} \langle \!\langle \Vii (x)\Vii (y)\rangle\!\rangle =
{\ii^2\over (x-y)^2 - \ii^2}.  \ee
The interaction between two dipoles depends both on the distance and
on the direction.  If $\ii=|\ii|i$, then the force between two dipoles
is repulsive if they are spaced horizontally and attractive if they
are spaced vertically. As the interaction rapidly decreases at large distances, one can
compute the thermodynamics of the dipole gas by performing  
Mayer (cumulant) expansion. The poles of the  pair-wise interaction potential 
at   $x-y=\pm\ii$ lead to a phenomenon called in \cite{Nekrasov:2009aa}
 clustering of instanton particles. The fundamental dipole can form `bound states' of 
 $n$ fundamental dipoles, whose field-theoretical counterpart are the exponential fields \re{compositep}. A composite particle made of $n$ fundamental dipoles behaves as 
 a pair of positive and negative electric charges spaced at distance $n\ii$.
 
 By the operator representation \re{dilogoper},  the $\caA$-functional 
 is the grand partition function of a non-ideal gas made of the fundamental particles and
 all kinds of composite particles.  The particles of this gas interact with the  external potential $\Phi(x)$ and pairwise as
\be \la{vavAmn} 
\langle \!\langle \CV_{m\ii}(x)\  \CV_{n\ii} (y)\rangle\!\rangle =
{\ii^2 \,  mn \over  (x-y+m\ii)(x-y - n\ii)}.  \ee
The effective one-dimensional theory \re{defYY}  describes the 
limit when only the dipole charge is taken into account, while the 
quadruple etc. charges, small by powers of $\ii$, are neglected.
The first term in our final formula \re{leadsubleadQFT} 
 corresponds to
the dilute gas approximation, in which the charges interact only with
the external potential,  while the sub-leading second term takes into
account the pairwise interactions.

  \section*{Acknowledgments}
  
 I.K. thanks J.-E. Bourgine, D. Fioravanti,  N. Gromov, S. Komatsu, Y. Matsuo and  F. Ravanini 
  for illuminating discussions.  We are grateful for the
 hospitality at the Simons Center for Geometry and Physics, where this
 work has been initiated.  EB is grateful for the hospitality at the
 University of Cologne.  This work has been supported by European
 Programme IRSES UNIFY (Grant No 269217), the Israel Science
 Foundation (Grant No.  852/11) and by the Binational Science
 Foundation (Grant No.  2010345).

 \footnotesize
%%% 
% \bibliography{/Users/vani/Files/PAPERS/PAPERSLIBRARY/ABib}

\begin{thebibliography}{10}

\bibitem{PhysRevLett.74.816}
B.~Sutherland, ``Low-Lying Eigenstates of the One-Dimensional Heisenberg
  Ferromagnet for any Magnetization and Momentum,'' {\em Phys. Rev. Lett.} {\bf
  74} (Jan, 1995) 816--819.

\bibitem{Babelon:Bernard:Smirnov:Quantization:Solitons}
O.~{Babelon}, D.~{Bernard}, and F.~A. {Smirnov}, ``{Quantization of Solitons
  and the Restricted Sine-Gordon Model},'' {\em Communications in Mathematical
  Physics} {\bf 182} (Dec., 1996) 319--354,
  \href{http://arXiv.org/abs/hep-th/9603010}{{\tt hep-th/9603010}}.

\bibitem{2011PhRvB..84v4503G}
G.~{Gorohovsky} and E.~{Bettelheim}, ``{Exact expectation values within
  Richardson's approach for the pairing Hamiltonian in a macroscopic system},''
  {\em Phys. Rev. B} {\bf 84} (Dec., 2011) 224503,
  \href{http://arXiv.org/abs/1111.1519}{{\tt 1111.1519}}.

\bibitem{Beisert:2003xu}
N.~Beisert, J.~A. Minahan, M.~Staudacher, and K.~Zarembo, ``{Stringing spins
  and spinning strings},'' {\em JHEP} {\bf 09} (2003) 010,
\href{http://arXiv.org/abs/hep-th/0306139}{{\tt hep-th/0306139}}.
%%CITATION = HEP-TH/0306139;%%.

\bibitem{Kazakov:2004qf}
V.~Kazakov, A.~Marshakov, J.~A. Minahan, and K.~Zarembo, ``{Classical / quantum
  integrability in AdS/CFT},'' {\em JHEP} {\bf 05} (2004) 024,
\href{http://arXiv.org/abs/hep-th/0402207}{{\tt hep-th/0402207}}.
%%CITATION = HEP-TH/0402207;%%.

\bibitem{Beisert-Rev}
N.~{Beisert {\it et al}}, ``{Review of AdS/CFT Integrability: An Overview},''
  {\em Letters in Mathematical Physics} {\bf 99} (Jan., 2012) 3--32,
  \href{http://arXiv.org/abs/1012.3982}{{\tt 1012.3982}}.

\bibitem{EGSV}
J.~{Escobedo}, N.~{Gromov}, A.~{Sever}, and P.~{Vieira}, ``{Tailoring
  three-point functions and integrability},'' {\em JHEP} {\bf 9} (Sept., 2011)
  28, \href{http://arXiv.org/abs/1012.2475}{{\tt 1012.2475}}.

\bibitem{Foda:3ptdeterminant}
O.~{Foda}, ``{$\mathcal{N}=4$ SYM structure constants as determinants},'' {\em
  Journal of High Energy Physics} {\bf 3} (Mar., 2012) 96,
  \href{http://arXiv.org/abs/1111.4663}{{\tt 1111.4663}}.

\bibitem{GSV}
N.~{Gromov}, A.~{Sever}, and P.~{Vieira}, ``{Tailoring Three-Point Functions
  and Integrability III. Classical Tunneling},'' {\em ArXiv e-prints} (Nov.,
  2011) \href{http://arXiv.org/abs/1111.2349}{{\tt 1111.2349}}.

\bibitem{3pf-prl}
I.~{Kostov}, ``{Classical Limit of the Three-Point Function of N=4
  Supersymmetric Yang-Mills Theory from Integrability},'' {\em Physical Review
  Letters} {\bf 108} (June, 2012) 261604,
  \href{http://arXiv.org/abs/1203.6180}{{\tt 1203.6180}}.

\bibitem{SL}
I.~{Kostov}, ``{Three-point function of semiclassical states at weak
  coupling},'' {\em Journal of Physics A Mathematical General} {\bf 45} (Dec.,
  2012) 4018, \href{http://arXiv.org/abs/1205.4412}{{\tt 1205.4412}}.

\bibitem{NSlavnov1}
N.~A. Slavnov, ``Calculation of scalar products of wave functions and form
  factors in the framework of the algebraic Bethe ansatz,'' {\em Theoretical
  and Mathematical Physics} {\bf 79} (1989) 502--508. 10.1007/BF01016531.

\bibitem{sz}
I.~{Kostov} and Y.~{Matsuo}, ``{Inner products of Bethe states as partial
  domain wall partition functions},'' {\em JHEP10(2012)168} (July, 2012)
  \href{http://arXiv.org/abs/1207.2562}{{\tt 1207.2562}}.

\bibitem{Nekrasov:2009aa}
N.~A. Nekrasov and S.~L. Shatashvili, ``Quantization of Integrable Systems and
  Four Dimensional Gauge Theories,'' \href{http://arXiv.org/abs/0908.4052}{{\tt
  0908.4052}}.

\bibitem{korepin-DWBC}
V.~E. Korepin, ``Calculation of norms of Bethe wave functions,'' {\em
  Communications in Mathematical Physics} {\bf 86} (1982) 391--418.
  10.1007/BF01212176.

\bibitem{Kazama:2013aa}
Y.~{Kazama}, S.~{Komatsu}, and T.~{Nishimura}, ``{A new integral representation
  for the scalar products of Bethe states for the XXX spin chain},'' {\em ArXiv
  e-prints} (Apr., 2013) \href{http://arXiv.org/abs/1304.5011}{{\tt
  1304.5011}}.

\bibitem{Muskelishvili}
N.~I. Muskhelishvili, {\em Singular Integral Equations: Boundary Problems of
  Function Theory and Their Application to Mathematical Physics}.
\newblock P. Noordhoff, 1953.

\bibitem{Mayer-MY}
C.~{Meneghelli} and G.~{Yang}, ``{Mayer-Cluster Expansion of Instanton
  Partition Functions and Thermodynamic Bethe Ansatz},'' {\em ArXiv e-prints}
  (Dec., 2013) \href{http://arXiv.org/abs/1312.4537}{{\tt 1312.4537}}.

\bibitem{JEB-Mayer}
J.-E. Bourgine, ``{Confinement and Mayer cluster expansions},''
\href{http://arXiv.org/abs/1402.1626}{{\tt 1402.1626}}.
%%CITATION = ARXIV:1402.1626;%%.

\bibitem{JimboMiwa-tau}
M.~Jimbo and T.~Miwa, ``Solitons and infinite dimensional Lie algebras,'' {\em
  Publ. RIMS, Kyoto Univ} {\bf 19} (1983) 943--1001.

\bibitem{G.Moore:2000aa}
G.Moore, N.Nekrasov, and S.Shatashvili, ``Integrating Over Higgs Branches,''
  {\em Commun.Math.Phys.} {\bf 209} (2000) 97--121,
  \href{http://arXiv.org/abs/hep-th/9712241}{{\tt hep-th/9712241}}.

\bibitem{Destri:1994bv}
C.~Destri and H.~J. De~Vega, ``{Unified approach to thermodynamic Bethe Ansatz
  and finite size corrections for lattice models and field theories},'' {\em
  Nucl. Phys.} {\bf B438} (1995) 413--454,
\href{http://arXiv.org/abs/hep-th/9407117}{{\tt hep-th/9407117}}.
%%CITATION = HEP-TH/9407117;%%.

\bibitem{0305-4470-24-13-025}
A.~Klumper, M.~T. Batchelor, and P.~A. Pearce, ``Central charges of the 6- and
  19-vertex models with twisted boundary conditions,'' {\em Journal of Physics
  A: Mathematical and General} {\bf 24} (1991), no.~13, 3111.

\bibitem{Feverati:2006tg}
G.~Feverati, D.~Fioravanti, P.~Grinza, and M.~Rossi, ``{On the finite size
  corrections of anti-ferromagnetic anomalous dimensions in N = 4 SYM},'' {\em
  JHEP} {\bf 05} (2006) 068,
\href{http://arXiv.org/abs/hep-th/0602189}{{\tt hep-th/0602189}}.
%%CITATION = HEP-TH/0602189;%%.

\bibitem{Bargheer:2008kj}
T.~Bargheer, N.~Beisert, and N.~Gromov, ``{Quantum Stability for the Heisenberg
  Ferromagnet},''
\href{http://arXiv.org/abs/0804.0324}{{\tt 0804.0324}}.
%%CITATION = 0804.0324;%%.

\end{thebibliography}
%  \bibliographystyle{/Users/vani/Files/PAPERS/PAPERSLIBRARY/utcaps}
%
 
\providecommand{\href}[2]{#2}\begingroup\raggedright\endgroup

\end{document}